\def\msol{{\rm M}_{\odot}}

\documentclass{aa}
\usepackage{graphicx}
\begin{document}
%
%
%
\newcommand{\ddp}[2]{\frac{\partial #1}{\partial #2}}
\newcommand{\dif}[1]{\ensuremath{{\rm d}#1}}

\title{Velocity-induced collapses of stable neutron stars}
\author{J. Novak}
\institute{
D\'epartement d'Astrophysique Relativiste et de Cosmologie
-- UMR 8629 du CNRS, Observatoire de Paris, F-92195 Meudon Cedex,
France\\
{\em Jerome.Novak@obspm.fr}}
\offprints{J. Novak}
\date{Received/Accepted}
\abstract{
The collapse of spherical neutron stars is studied in General
 Relativity. The initial state is a stable neutron star to which an
 inward radial kinetic energy has been added through some velocity
 profile. For two different equations of state and two different
 shapes of velocity profiles, it is found that neutron stars can
 collapse to black holes for high enough inward velocities, provided
 that their masses are higher than some minimal 
 value, depending on the equation of state. For a
 polytropic equation of state of the form $p=K\rho^\gamma $, with
 $\gamma = 2$ it is found to be $1.16 \left( \frac{K}{0.1}
 \right)^{0.5} \msol$, whereas for a more realistic one (described in
 \cite{PonREPL00}), it reads $0.36 \msol $. In some cases of collapse
 forming a black hole, part of the matter composing the initial
 neutron star can be ejected through a shock, leaving 
 only a fraction of the initial mass to form a black hole. Therefore,
 black holes of very small masses can be formed and, in particular,
 the mass scaling relation, as a function of initial velocity, takes
 the form discovered by \cite{Cho93} for critical collapses.
}

\titlerunning{ Velocity-induced collapses}
\authorrunning{J.~Novak}
\maketitle

\section{Introduction}
%
In the early eighties \cite{ShaT80} addressed the
question of formation of black holes in astrophysical collapses. Among
other, they asked the (theoretical) question of the minimum mass of a
black hole formed by the adiabatic collapse of a stellar core and ``In
particular, can the effective mass-energy potential barrier associated
with equilibrium configurations be penetrated by low-mass cores with
substantial inward, radial kinetic energy?''. In astrophysical
scenarios, black holes can also form from accretion-induced collapses
of neutron star, if the neutron star is part of a binary system or
during a {\it supernova} event, when part of the envelop fails to
reach escape velocity and falls back onto the new born neutron
star. Therefore, one can also ask the question: how much of inward
kinetic energy has to be put to a neutron star to make it collapse to
a black hole? Can a neutron star always collapse to a black hole,
provided that it gets enough kinetic energy?

If one looks at {\em static} neutron
star models, these are stable against perturbations if their masses are
lower than some maximal mass and the density is also lower than the
``critical'' density corresponding to this maximal mass (they are
located on the so-called {\em stable branch}). They can collapse to
form a black hole 
when the central density is higher than the critical one (see
e.g. \cite{Gou91}). The question is then 
whether {\em stable} neutron stars (i.e. with lower density and mass than the
critical ones) can be ``pushed'', with a certain amount of inward
kinetic energy, to form a black hole, and what is the minimal mass of
the formed black hole and/or progenitor? The problem of the
minimal mass of a black hole has been solved, first from a
more mathematical point of view, by \cite{Cho93},
who discovered 
the ``critical collapse'' phenomena in General Relativity (see also
Sec.~\ref{ss:link} for a discussion). Still, the problem of a stellar
core/neutron star with inward kinetic energy have never been
completely studied; to our knowledge, only a partial study was done by
\cite{Gou92}.  
The aim of this paper is to numerically follow the collapse of 
stable neutron stars, with inward velocity profile, and determine the
initial conditions necessary to obtain a black hole as result of the
collapse; the purpose is also to allow for a realistic equation of
state (EOS), to study dependence on the EOS, as well as on the initial
velocity profile, giving kinetic energy. Therefore, emphasis shall be
put on neutron star properties in General Relativity and the question
of the physical mechanism giving this kinetic energy will not be
addressed. 

The complete model in General Relativity is described in
Sec.~\ref{s:model}, including the system of partial derivative
equations (\ref{ss:equa}), the two EOS used (\ref{ss:eos}), and the
procedure for obtaining initial numerical models (\ref{ss:IC}). Time
evolution is studied in Sec.~\ref{s:evolution}, where two different
possible evolutions for neutron stars are given (\ref{ss:smooth} and
\ref{ss:shock}). Numerical results are displayed in
Sec.~\ref{s:resu}: together with a link with the ``critical collapse''
paradigm (\ref{ss:link}), properties of initial data (\ref{ss:mnrj})
and dependence on the parameters (\ref{ss:param}) are discussed.
Finally, Sec.~\ref{s:conc} summarizes the results and gives some
concluding remarks.

\section{Evolution of spherically symmetric neutron stars}\label{s:model}
\subsection{Field and Matter equations}\label{ss:equa}
The equations for the evolution of the matter and gravitational fields
are derived from the Einstein equations using the simplifying
assumption of spherical symmetry. Space-time
is foliated into spacelike hypersurfaces
$\Sigma_t $
and equation are written in terms of the classical 3+1 formalism of
General Relativity 
(see e.g. \cite{ArnDM62}). The real parameter $t$ is called the {\em coordinate
time} and it can be shown that, making the choice of {\em polar time
slicing} and {\em radial gauge}, in spherical symmetry the metric can
be expressed in the form (generalization of the Schwarzschild metric,
see also \cite{Gou91}):
\begin{eqnarray}
g_{\mu\nu}\dif{x^\mu}\dif{x^\nu} &=& \nonumber \\
 -N^2(r,t)\dif{t^2} +
A^2(r,t)\dif{r^2} &+& r^2(\dif{\theta^2} + \sin^2\theta \dif{\phi^2}),
\label{e:metrique}
\end{eqnarray}
where $N(r,t)$ is called the {\em lapse function}. The two functions
$N$ and $A$ will often be replaced by $\nu(r,t)$ and $m(r,t)$, defined
as\footnote{We use geometrized units in which the speed of light $c$
and Newton's gravitational constant $G$ are equal to unity}:
\begin{equation}
\nu = \ln (N)\ {\rm and}\ A = \left( 1 - \frac{2m}{r} \right)^{-1/2}.
\label{e:defnum}
\end{equation}
The 4-velocity of the fluid is denoted $v^\mu$, the fluid radial {\em
coordinate} velocity being thus:
$$
\frac{\dif{r}}{\dif{t}} = \frac{v^r}{v^0};
$$
and the ``physical'' fluid radial velocity $U$, as measured locally by
the hypersurfaces observer is defined by:
\begin{equation}
U = \frac{A}{N} \frac{\dif{r}}{\dif{t}}.
\end{equation}

Following \cite{RomIMM96}, the stress-energy tensor was taken to be
that of a perfect fluid; so hydrodynamical equations in General
Relativity ($\nabla_\mu T^\mu_\nu$ = 0) have been written
in the form of a system of conservation laws 
\begin{equation}
\ddp{\vec{u}}{t} + \frac{1}{r^2}\ddp{}{r}\left[ r^2\frac{N}{A}
\vec{f}(\vec{u}) \right] = \vec{s}(\vec{u}), \label{e:conshyp} 
\end{equation}
where $\vec{u} = \{D, \mu, \tau \}$ is the vector of evolved
quantities, $\vec{f}(\vec{u})$ the vector of fluxes and $\vec{s}(\vec{u})$
the vector of sources (for details see \cite{RomIMM96}). The evolved
quantities are defined from hydrodynamical variables, baryon 
and total energy densities in the fluid frame ($n_B$ and $e$) by:
\begin{eqnarray}
D &=& A\Gamma n_B, \nonumber \\
\mu &=& (E+p)U, \label{e:defqhyd}\\
\tau &=& E - D, \nonumber 
\end{eqnarray}
where $p$ is the fluid pressure given by the EOS (see \ref{ss:eos}),
$\Gamma = (1-U^2)^{-1/2}$ is the Lorentz factor of the fluid, and $E=
\Gamma^2(e+p)-p$.  

The gravitational field equations in radial gauge, polar slicing and
spherical symmetry reduce to two equations (with no time
evolution of the gravitational field):
\begin{eqnarray}
\ddp{m}{r} &=& 4\pi r^2 E, \label{e:dmdr}\\
\ddp{\nu}{r} &=& A^2\left(\frac{m}{r^2} + 4\pi r(p+(E+p)U^2) \right).
\label{e:dnudr} 
\end{eqnarray}
The system therefore consists of three evolution equations for the
hydrodynamical variables (Eqs.~\ref{e:conshyp}) and two constraint
equations for the gravitational field
(Eqs.~(\ref{e:dmdr})--(\ref{e:dnudr})). Finally, an equation of state
$p(n_B,e)$ closes the system.

\subsection{Equation of state}\label{ss:eos}

In this work two different EOS for nuclear matter have
been used, to describe microscopic properties of neutron stars. Apart
from the initial condition models, which were calculated with an
equation of state $p(n_B)$ for cold, catalyzed matter, the used
equations of state were of the form $p(n_B,e)$. The first EOS used is
the well-known {\bf ideal gas} model:
\begin{equation}
p = (\gamma -1)(e - m_B n_B) 
\end{equation}
where $m_B=1.66\times 10^{-27}$ kg is the baryon mass and $\gamma$ is
the index. We chose $\gamma =2$ which may mimic relatively
well the properties of neutron star matter but always keeps a sound
speed lower than $c$ ({\em causal} EOS). This EOS has been used not
only because it is very convenient from a numerical point of view, but
also because it allows for the calibration and test of the code. For
obtaining initial conditions, the following expression has been used 
\begin{equation}
p(n_B) = K n_0 m_B \left( \frac{n_B}{n_0} \right)^\gamma ;
\label{e:eospoly}
\end{equation}
with $K=0.1$ (as in the study of rotating neutron star models by
\cite{SalBGH94} ) and 
\begin{equation}
n_0 = 0.1 {\rm fm}^{-3}. \label{e:defn0}
\end{equation}

To get a more realistic EOS for nuclear matter interactions, we used
the EOS described in \cite{PonREPL00}. It is a relativistic field
theory model supplemented by nonlinear scalar self
interactions. Nucleons $(n,p)$ interact via the exchange of $\sigma-$,
$\omega-$ and $\rho-$mesons. The contribution from leptons is given by
its non-interacting form, since their interactions give negligible
contributions. Moreover, the star being at chemical equilibrium with
respect to the weak processes, the neutrino chemical potentials are
zero. However, as a difference from \cite{PonREPL00}, kaon
interactions are not considered in this work. The EOS was tabulated
and, during the time integration, the interpolation has been done
using bi-cubic splines. This may not be as precise as the method
presented by \cite{Swe96} (using bi-{\em quintic} interpolation),
but thermodynamical consistency is still preserved. At low densities,
this EOS has been smoothly joined with a polytrope. As far as initial
conditions are concerned, the same model has been used, but the
temperature has been set to zero.

\subsection{Initial Conditions} \label{ss:IC}
%

To get initial numerical conditions, we first obtained stable
spherical neutron star models and then, to add a velocity profile. A
{\em stable} neutron star is defined by the fact that its central
density is lower than the critical one ($n_B^{\rm crit}$), defined by
($M_g$ being the total gravitational mass of the star {\bf and $n^c_B
= n_B(r=0)$}):
\begin{equation}
\left. \frac{\dif{M_g}}{\dif{n^c_B}} \right|_{n^c_B = n_B^{\rm crit}} =
0. \label{e:nbcrit}
\end{equation}
The stable model is easily computed, integrating the well known
Tolman-Oppenheimer-Volkoff (TOV) system, which is the static (all
$\partial / \partial t$ terms and $U$ are set to zero) limit of
momentum evolution equation in system (\ref{e:conshyp}), plus the
equations for gravitational fields
Eqs.~(\ref{e:dmdr})--(\ref{e:dnudr}). For the polytropic equation of state
($\gamma = 2, K = 0.1$), one has 
$n_B^{\rm crit}= 3.18$ $n_0$ and the maximal stable mass is $M_g
= 3.16 \msol$. For the tabulated EOS of \cite{PonREPL00} (see previous
section), one has $n_B^{\rm crit}= 10.5$ $n_0$ and the maximal stable
mass is $M_g = 2.08 \msol$. The rather high value for maximal mass
for the $\gamma = 2$ polytropic EOS is linked to an also high value
chosen for $K$. Thus mass scales between both EOS used in this work
were different. {\bf However for polytropes of the form
(\ref{e:eospoly}) with $\gamma = 1+1/n$, the constant $K^{n/2}$ has
units of length in geometrized units. Following \cite{CooST94}, one
can use this constant to set the fundamental length scale of the
system. Defining dimensionless quantities as in \cite{CooST94}
($\bar{t} = K^{-n/2}t$, $\bar{p} = K^np$, $\bar{n}_B = K^n n_B$, ...),
one can see that in our case all masses (even in dynamical evolution)}
scale like $K^{0.5}$. 

Once the density and gravitational fields are computed, the inward
velocity profile is added in the r.h.s. of gravitational field
equations (\ref{e:dmdr})--(\ref{e:dnudr}). The metric coefficients are then
relaxed in order to take into account the contribution from the
kinetic energy in the total gravitational mass. Therefore, initial
conditions are consistent with the gravitational field equations.
The considered profiles were of the form
$$
U(r) = \frac{A(r)}{N(r)}V\left(\frac{r}{R_{\rm surface}}\right),
$$
where $V(x)$ has one of these forms ($x=r/R_{\rm surface}$):
\begin{eqnarray}
V(x) &=& \frac{V_{\rm amp}}{2} (x^3 - 3x) \label{e:V1(X)}\\
V(x) &=& \frac{27V_{\rm amp}}{10\sqrt{5}} \left(x^3 - \frac{5x}{3}\right) \label{e:V2(X)}
\end{eqnarray}
($V_{\rm amp}$ is the parameter defining the amplitude of the
profile).

Both are such that the minimal value in [0,1] is $-V_{\rm amp}$, but
(\ref{e:V1(X)}) 
verifies $V'(1)=0$ and (\ref{e:V2(X)}) has a null divergence at the
surface. These profiles correspond to what is usually observed in
collapses of neutron stars (see e.g. \cite{Gou91}) or in the formation
of neutron stars (see e.g. \cite{RomIMM96}).
The initial minimal value of the velocity $U(r)$ in the interior of the star
($r \in [0,R_{\rm surface}]$) is noted $-U_{\rm min}$. It shall also be
called the initial velocity profile amplitude.
\section{Dynamical scenarios}\label{s:evolution}

Numerical time integration has been done using High Resolution
Shock-Capturing schemes (HRSC, see \cite{BanFIMM97}) for the
hydrodynamical system (\ref{e:conshyp}). The metric constraint
equations (\ref{e:dmdr})--(\ref{e:dnudr}) were integrated using standard
finite-differences methods. The need of using numerical methods able to
handle shocks comes from the fact that strong discontinuities can form
(see Sec.~(\ref{ss:shock})), as it has first been observed by
\cite{Gou92}.
Unfortunately, this last
study has been limited by the use of {\em spectral methods}, unable to
handle shocks. 

\subsection{Direct collapse}\label{ss:smooth}

For a stable star close to the maximal mass, but with an inward
velocity profile, there may be two final issues. If the velocity is
relatively small in amplitude, the star enters an (theoretically)
infinite series of oscillations, for no viscous nor radiative damping
is present. With higher velocity, a collapse to a black hole occurs
``normally'', almost like for an unstable neutron star configuration
(as in e.g.~\cite{Gou91}). 
\begin{figure}
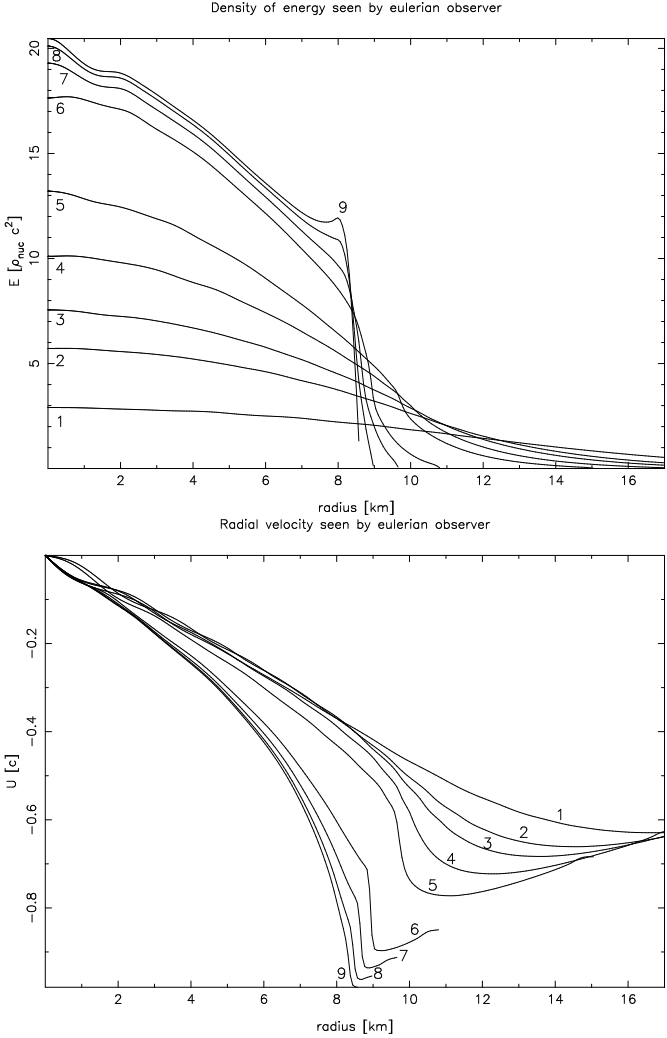

\resizebox{\hsize}{!}{\includegraphics[angle=-90]{ener26.ps}}
\resizebox{\hsize}{!}{\includegraphics[angle=-90]{vit26.ps}}
\caption{ {\bf Snapshots of eulerian energy density $E$} and radial 
velocity profiles $U$, for a 2.74~$\msol$ neutron star described by a
$\gamma =2$ polytropic EOS, with an initial velocity profile of type
(\ref{e:V2(X)}), and a starting minimal value of the velocity of
${\mathbf -}U_{\rm min} =  {\mathbf -}0.6 c$. {\bf Labels correspond to
the following times: $t_1=0.054$ ms, $t_2=0.088$ ms, $t_3=0.104$ ms,
$t_4=0.126$ ms, $t_5=0.149$ ms, $t_6=0.202$ ms, $t_7=0.225$ms,
$t_8=0.248$ ms and $t_9=0.271$ ms. All profiles have been cut at
the surface of the star, defined by the surface density of the initial
configuration.}} 
\label{f:smoo}
\end{figure}
Figure~\ref{f:smoo} shows {\bf energy ($E$)} and velocity profiles
{\bf at different moments of} the
collapse of such a neutron star. It corresponds to an initial
configuration of 2.74 $\msol$, where the initial velocity profile is
of the type (\ref{e:V2(X)}), with $V{\rm amp} = 0.4 c$, which gives an
initial minimal radial velocity of ${\mathbf{-}}U_{\rm min} = 
{\mathbf{-}}0.6 c$. 
This collapse has been followed up to $t=0.48$ ms, when then central
value of the lapse became $5.13 \times 10^{-5}$, with a radial
velocity at the outer edge of -0.99 $c$. Due to the choice of polar
slicing which avoids the appearance of singularities, 
the horizon of a black hole cannot be numerically described. However, 
the final state is a black hole, but from the ``frozen star''
viewpoint, with a collapse of the lapse. These criteria have been used
in
Sec.~\ref{s:resu} to determine whether the result of the collapse
was a black hole or not, namely a central value of the lapse $N(r=0) <
10^{-4}$ with an ingoing radial velocity. If both of these conditions
are fulfilled, it seems unlikely that the collapse lead to anything
else than a black hole.

Still, a difference from \cite{Gou91} in the dynamical evolution can be seen in
Fig.~\ref{f:smoo}: a strong gradient appears in the velocity profile
around $r=9$ km, even quite before the ``frozen star'' regime (where the
metric potentials $A$ and $N$ also exhibit strong gradients). This had
already been observed by \cite{Gou92}, where it limited the study. It
is shown in Sec.~\ref{ss:shock} that, with higher initial velocities,
a shock can form and part of the infalling matter can be
ejected. In the present ``direct collapse'' regime all the matter
ends in the black hole so that the mass of the resulting object is that
of the initial neutron star. In Sec.~\ref{s:resu}, this has been used
to distinguish between both possible types of collapses.

\subsection{Shock and bounce}\label{ss:shock}

For some initial velocities, the ``strong gradient''
of previous section turns into a shock, and part of the infalling
material is ejected. 
\begin{figure}
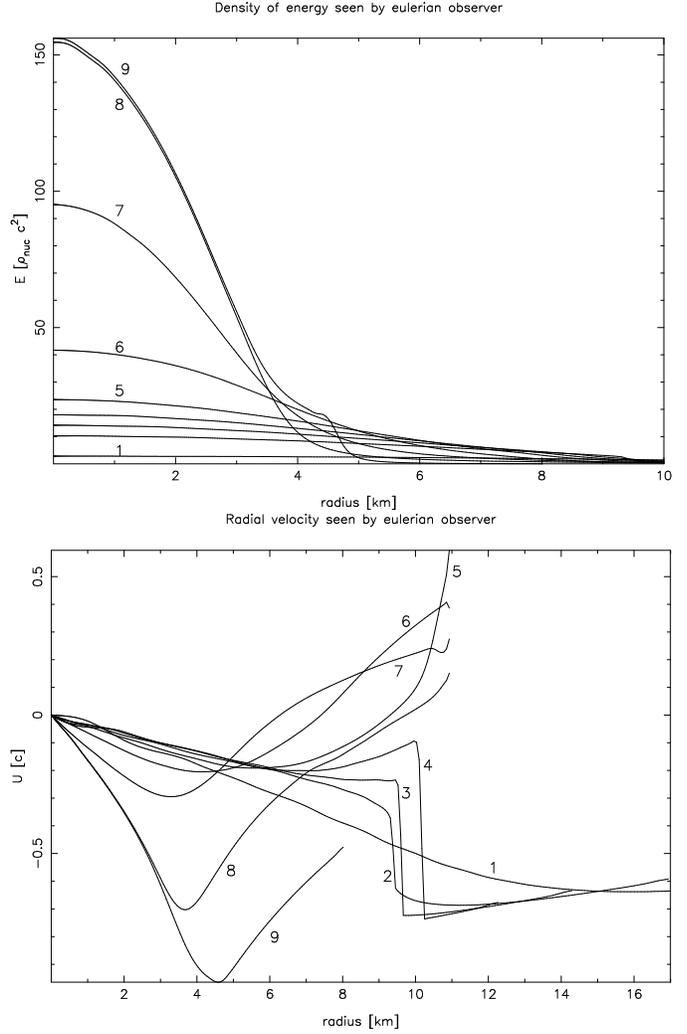

\resizebox{\hsize}{!}{\includegraphics[angle=-90]{ener14.ps}}
\resizebox{\hsize}{!}{\includegraphics[angle=-90]{vit14.ps}}
\caption{ {\bf Eulerian energy} density and radial velocity profiles, for a
polytropic $\gamma =2$ 1.91~$\msol$ neutron star {\bf collapse}, with
an initial velocity profile of type (\ref{e:V2(X)}), and a starting
minimal value of the velocity of $-0.62 c$. The displayed profiles
correspond to {\bf the following times: $t_1=0.064$ ms, $t_2=0.123$
ms, $t_3=0.143$ ms, $t_4=0.16$ ms, $t_5=0.178$ ms, $t_6=0.216$ ms,
$t_7=0.265$ ms, $t_8=0.348$ ms and $t_9=0.477$ ms. All profiles have
been cut at the surface of the star, defined by the surface density
of the initial configuration.}} 
\label{f:shock}
\end{figure}
This can be seen on Fig.~\ref{f:shock} where {\bf energy} density and radial
velocity profiles are displayed, for an initial neutron star of 1.91
$\msol$ (the mass is lower than that of Fig.~\ref{f:smoo} due to a
lower initial central baryon density). The shock appears {\bf around}
$t=t_3$,  $r=10$ km
and then moves out of the numerical grid ($t_4$) due to the
accumulation of matter still falling at $r>10$ km. The velocity on the
left side of the discontinuity grows and reaches positive values.
Later during the
collapse, one can see that part of the falling matter starts moving
outward and reaches escape velocity from the central object $t_5$. This of
course reduces the possibility of forming a black hole, since matter
is spatially less concentrated. Still, the dynamics of the central
region (where the velocity is still inward) can proceed to a black
hole, which shall then accrete matter which has not reached escape
velocity. This has been observed in the collapse displayed on
Figs.~\ref{f:shock}  where, at the end of the
computation ($t=0.63$ ms), the central
lapse was $N(r=0) = 1.18\times 10^{-9}$ and the velocity on the edge
of the ``frozen star'' equal to -0.999 $c$. The central region would
collapse ``directly'', as described before in
Sec.~\ref{ss:smooth}. The final mass of the black hole 
was 1.61 $\msol$ (and the areal radius in RGPS coordinates equal to 4.77 km),
which shows that almost 16\% of the initial matter of the neutron
star has been ejected.

\section{Numerical results}\label{s:resu}
The dynamical fate of neutron stars with initial velocity are now
studied, keeping in mind the different scenarios described in previous
section and varying the mass of the progenitor, as well as the
amplitude of the initial velocity profile.  This exploration of the
parameter space shows the possibility of formation of very low mass black
holes, which can be seen as a feature of the ``critical collapse''
phenomena. 

\subsection{Link with critical collapses}\label{ss:link}

The critical collapse phenomenon (for an interesting review see
\cite{Gun99}) was discovered in the early 90s by
\cite{Cho93}, who was numerically studying the gravitational collapse
of spherically symmetric massless scalar field. Depending on some
parameter of the initial conditions (which is generically noted
$p$), the final result of the collapse would be a black hole ($p$
large) or the dispersion of the field ($p$ small). He discovered that,
when fine-tuning this parameter, he could get black holes of
arbitrarily small masses. Moreover, the relation giving the mass of
the resulting black hole, close to $p_*$ the minimal value to form a
black hole, appeared to be universal in the form:
\begin{equation}
M_{BH} \simeq C(p - p_*)^\alpha, \label{e:relcrit}
\end{equation}
where $C$ is a constant and $\alpha$ is called the {\em critical
exponent}. The space-time obtained with $p=p_*$ shows the very
interesting geometrical property of {\em self-similarity}. 

There have been many works on critical collapses
since this pioneering one and, in particular, the study concerning
perfect fluids by \cite{NeiC00} is of particular interest for our
study. It is shown that perfect fluid collapses, in
ultra-relativistic regime, also exhibit ``critical'' behavior. Our
work can easily be connected to this one, if one considers, at fixed
central density of the initial neutron star, the parameter $p$ to be
e.g. the parameter $V_{\rm amp}$ or the amplitude of the initial
velocity profile  $U_{\rm min}$. One can 
then see that, by fine-tuning the parameter $V_{\rm amp}$ ($U_{\rm
min}$ being a monotonic and continuous function of $V_{\rm amp}$) of
(\ref{e:V1(X)}) or (\ref{e:V2(X)}), one could get black holes of
arbitrarily small masses, as a result of the velocity-induced
collapse. We have not investigated the domain of very small
black hole masses, for our code was not designed for it. Still, 
starting with a $1.16 \msol$ neutron star and an initial velocity
profile given by (\ref{e:V2(X)}) (and $V_{\rm amp} = 0.79732$), the
result of the evolution was a $3.7 \times 10^{-2} \msol$ black
hole. This corresponds to the lowest black hole mass range (about $10^{-2}
\msol$) obtained with our code. In order to test the code against
the relation (\ref{e:relcrit}), the central density of the initial
configuration has been kept constant, and only the parameter $V_{\rm
amp}$ of (\ref{e:V2(X)}) has been varied (and thus
$U_{\rm min}$). 
\begin{figure}
\resizebox{\hsize}{!}{\includegraphics[angle=-90]{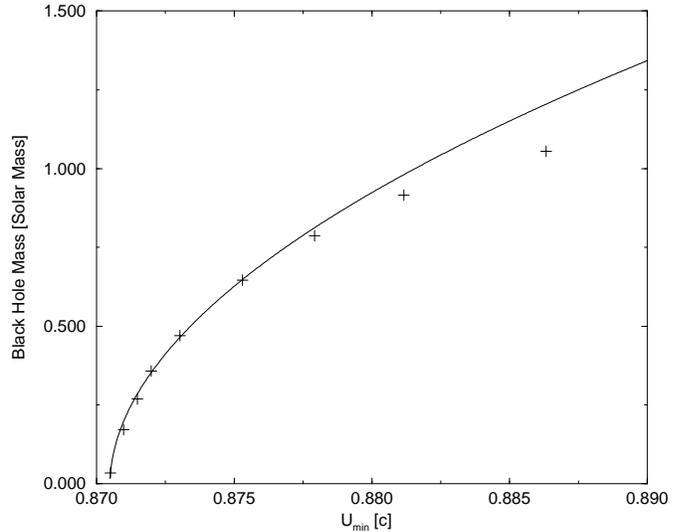}}
\caption{Masses of the black holes ($+$) formed through velocity-induced
collapses, for a polytropic EOS and with an initial velocity profile
 of type (\ref{e:V2(X)}). The initial central density is $0.1 n_0$
 (see Sec.~\ref{ss:eos}) for all the runs. These masses have been
 fitted  by a formula of type (\ref{e:relcrit}) (solid line). }
 \label{f:mass2b}
\end{figure}

The results displayed on Fig.~\ref{f:mass2b} correspond to the masses
of black holes resulting from collapses, if the central density of the
initial configurations is held fixed to $0.1 n_0$. These masses are
drawn (crosses) as a function of the minimal value of the velocity $U$ inside
the neutron star. The relation (\ref{e:relcrit}) is verified, at least
for ``small'' masses and the best fit (shown on the figure in solid
line) corresponds to $\alpha = 0.52$, $C = 10.4 \msol$ and $p_* =
0.8705 c$ ($U_{\rm min} = p$). These values has shown to depend on
the particular central density used for the initial
configurations. The fit is poor for large masses and, in particular,
the relation (\ref{e:relcrit}) breaks down completely near the maximal
mass for stable static neutron stars (when central density reaches
$n_B^{\rm crit}$). {\bf The mass scaling exponent $\alpha = 0.52$ differs
from the result obtained by \cite{NeiC00} ($\alpha = 1$) who used an
ultrarelativistic EOS with $\gamma = 2$. Although the result from
\cite{NeiC00} is rather universal, the study here starts from very
different initial conditions (neutron star/exponential distribution of
matter in their case). In particular}, here, the maximal mass for
stable neutron 
stars gives a {\em mass limit} to the problem. {\bf Another difference is
that the parameter used to get the relation (\ref{e:relcrit}) is the
amplitude of a velocity profile, which has not been studied by
\cite{NeiC00}. Finally, it has to be stressed out that the family of
initial conditions used for this study depends on {\em two} parameters
(velocity amplitude and central density), the work of \cite{NeiC00}
assumed only a one-parameter family of initial data. The role played
by the central density as a parameter is not that of a ``critical''
one: if one considers the family of initial data given by static
neutron stars (without any velocity profile) of increasing central
densities, then the value $n_B^{\rm crit}$ defined by (\ref{e:nbcrit})
would seem to be the same as $p_*$ in controlling the formation of a
black hole, but there is no critical behavior at this point.}
Still it is interesting to note that the mass scaling relation is
valid for {\bf velocity induced} neutron star collapses, at least in
the limit of low masses (and therefore relatively low central
densities).

\subsection{Masses of progenitors - Initial velocities}\label{ss:mnrj}

Once the EOS and type of velocity profile are chosen, a neutron star
model serving as initial condition for velocity-induced collapse is
completely specified by two parameters: the central baryonic density
$n^c_B$ and the parameter $V_{\rm amp}$ or, equivalently, the
neutron star gravitational mass and the amplitude of the initial
velocity profile $U_{\rm min}$. For each point ($n^c_B$, $U_{\rm min}$)
in this parameter space the question of knowing whether the
corresponding neutron star would collapse to a black hole have been
addressed. First, it has been found that there exists a maximal
value for $U_{\rm min}$, for which an initial model of neutron star
could be computed. {\bf For higher velocities, it appeared to be
numerically impossible to compute the metric coefficients taking into
account this large amount of kinetic energy: the relaxation would fail
to converge, leading to 
velocity $U > c$. This may
come from the way these initial conditions are set: the addition
of a velocity profile to a static model does not give a well defined 
result. Matter is not at equilibrium, whereas the metric potentials
are static. Therefore an improvement of this study would be to
consider more realistic initial conditions (as for example in
\cite{HawC00}), where the velocity profile is not set {\it ad hoc} but
comes from a dynamical interaction.}

\begin{figure}
\resizebox{\hsize}{!}{\includegraphics[angle=-90]{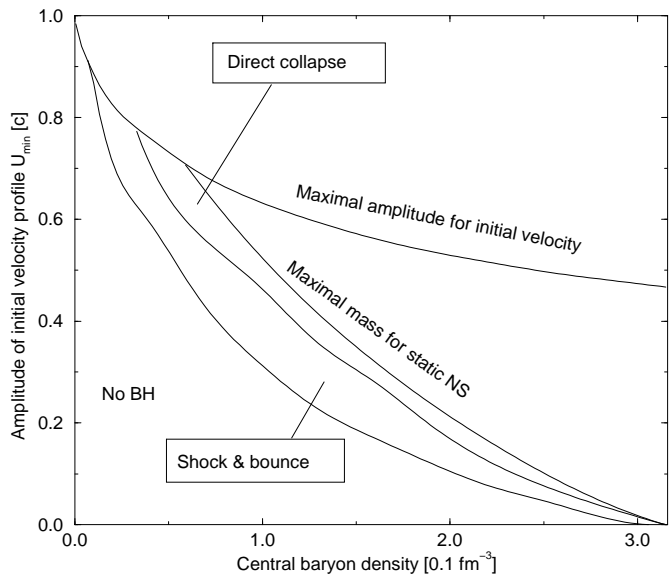}}
\caption{Parts of the parameter space plane ($n^c_B$, $U_{\rm min}$)
defining the fate of a neutron star with a polytropic EOS ($\gamma
 =2$) and an initial velocity profile defined by (\ref{e:V2(X)}).}
 \label{f:rhovpol}
\end{figure}

This maximal value of $U_{\rm min}$ depends on the central
density of the neutron star, as displayed on
Fig.~\ref{f:rhovpol}; one can see that for central densities going to
zero (and therefore also masses), this maximal value goes to one. This
figure also shows the various fates of a neutron
star in the ($n^c_B$,$U_{\rm min}$) plane, for the
($\gamma = 2$) polytropic EOS of Sec.~\ref{ss:eos} and an initial
velocity profile of type (\ref{e:V2(X)}).
Curves are described from top-right to bottom-left part of the figure. 
A neutron star set to a point of this parameter space can become a
black hole or not. Still, some region is not really interesting to
study.  
The second curve to be displayed selects neutron stars with a mass
lower than the maximal one, as defined in Sec.~\ref{ss:IC}. Studying
collapses for neutron stars with higher masses is not relevant since
one knows that in that case the neutron star shall end as a black
hole; the additional velocity profile of the neutron
star only gives kinetic energy so that the total mass of the star
become larger than the maximal one for stars at rest. It was then
observed that this collapse generally 
proceeds ``directly'', i.e. no shock is present. One more curve gives
the limit between the direct collapse domain and the part of the
parameter space for which the star undergoes a shock and bounce, as
described in Sec.~\ref{ss:shock}. Finally, if the parameters of the
initial neutron star model are below the last curve, the black hole
does not form, the matter being ejected to infinity. 
\begin{figure}
\resizebox{\hsize}{!}{\includegraphics[angle=-90]{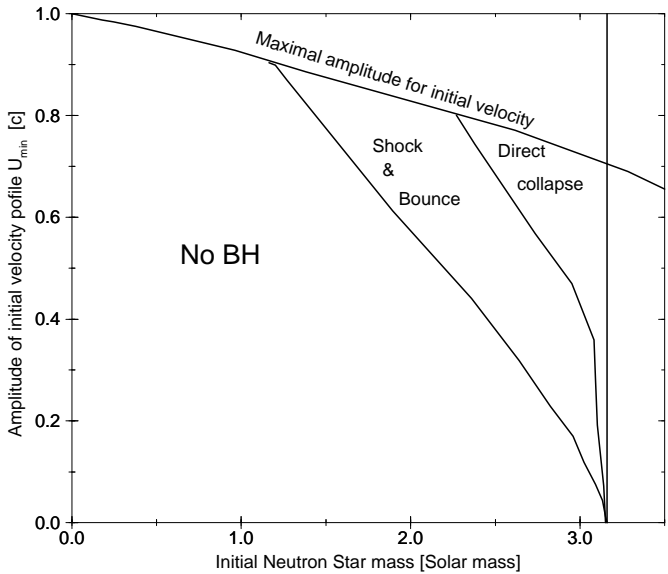}}
\caption{Parts of the parameter space plane ($n^c_B$, $M_g$)
defining the fate of a neutron star with a polytropic EOS ($\gamma
 =2$) and an initial velocity profile defined by (\ref{e:V2(X)}).}
 \label{f:mgvpol}
\end{figure}
The same regions are displayed in the $(M_g,U_{\rm min})$ plane, on
Fig.~\ref{f:mgvpol}. In particular, one can see that there exists a
minimal mass for the initial neutron star, to form a black hole. This
mass is, in the case of a polytrope $\gamma=2$ and an initial velocity
profile (\ref{e:V2(X)}):
\begin{equation}
M_{\rm min} = 1.164 \msol.
\end{equation}
This figure also shows the minimal neutron star mass for which direct
collapse (i.e. no ejection of matter) can occur, it found to be 2.276
$\msol$. 

\subsection{Dependence on EOS and velocity profiles}\label{ss:param}

Although the velocity profile (\ref{e:V2(X)}) is supposed to occur in
most of physical scenarios of gravitational collapse of compact stars,
one has to check the dependence of results 
of previous sections on a change of this velocity profile. Therefore
the study of Sec.~\ref{ss:mnrj} has been undertaken with the initial
velocity profile for neutron stars being given by (\ref{e:V1(X)}). For
initial configurations, one can see that this change in the velocity
profile induces, for given values of $n^c_B$ and $U_{\rm min}$, a change in the
distribution of kinetic energy in the star and thus a change in the
total gravitational mass $M_g$, with respect to the velocity
distribution given by (\ref{e:V2(X)}). Still, qualitative results
concerning collapses were retrieved: direct collapse or bounce as
dynamical scenarios and formation of low-mass black hole by fine-tuning
the parameter $U_{\rm min}$. The scaling relation (\ref{e:relcrit})
has been studied for a fixed central density (chosen to be $0.1 n_0$
as in Sec.~\ref{ss:link}) and the best fit of results gives the
relation:
$$
M_{BH} \simeq 6.98 \times (U_{\rm min} - 0.8991)^{0.52} \msol ,
$$
for which the exponent ($\alpha = 0.52$) is the same as that obtained using
velocity profile (\ref{e:V2(X)}). The two other coefficients (noted
$C$ and $p_*$ in Sec.~\ref{ss:link}) are different for the reasons
stated above, concerning the difference in kinetic energy
distribution. The borders of different regions of Fig.~\ref{f:mgvpol}
also slightly change, nevertheless we get as 
minimal mass for the initial neutron star to form a black hole for a
$\gamma = 2$ polytrope and an initial velocity profile (\ref{e:V1(X)}):
\begin{equation}
M_{\rm min} = 1.155 \msol.
\end{equation} 
This is less than 1\% different from the minimal mass for velocity profile
(\ref{e:V2(X)}), that is within global numerical error. One then may
suppose that this quantity is little dependent on the particular type
of velocity profile chosen.

The second global ``parameter'' one would like to change is the EOS
for nuclear matter, since a rather important uncertainty exists on
properties of neutron star matter. In particular, the $\gamma=2$
polytrope used in this study had the advantage of rapid computation as
well as numerical stability, combined with a rather good description
of neutron matter properties. But if one wants to go further, it is
then necessary to use a more realistic EOS, as for example the one
described in Sec.~\ref{ss:eos}. As stated in that section, maximal
mass for static neutron stars is found to be $M_g^{\rm crit} = 2.08
\msol$, for a corresponding central density of $n^c_B = 10.5
n_0$. {\bf Therefore, results from computations using this second EOS
would show lower masses and higher central densities than those using
the analytical EOS}. The parameter 
space ($n^c_B$, $U_{\rm min}$) has been studied, using the velocity
profile given by (\ref{e:V2(X)}). As for the $\gamma = 2$ polytrope,
neutron stars with realistic EOS would collapse to form a black hole,
provided that the amplitude of the velocity profile be large
enough. This collapse could occur directly, as in the case of {\em unstable}
neutron stars or with the bounce of a part of the matter, which would
be ejected, allowing only for the central region to form a black
hole. The mass scaling relation is recovered, for relatively small
initial central densities; for example, with $n^c_B = 1.5 n_0$, the
following relation is found:
$$
M_{BH} \simeq 5.79 \times (U_{\rm min} - 0.7519)^{0.71} \msol ,
$$
which has been tested for black holes in the mass range $5\times
10^{-3}-0.7 \msol$. The global topology of regions described in
figures~\ref{f:rhovpol} and \ref{f:mgvpol} is kept and the minimal
mass for a neutron star (with matter described by this realistic EOS)
to form a black hole is found to be 
\begin{equation}
M_{\rm min} = 0.36 \msol.
\end{equation} 
During the collapses leading to very low-mass black holes (a few
$10^{-2} \msol$), the central
regions of the star would reach very high densities and it has been
checked that sound velocity had never become higher than $c$.

\section{Summary  and conclusions}\label{s:conc}

With an important amount of inward kinetic energy, stable neutron stars may
collapse to a black hole, overcoming the potential barrier which
separates both types of objects. However, there seem to exist an
absolute lower mass limit, depending on the particular equation of
state used to describe neutron star matter properties, below which
neutron stars cannot pass this barrier. This lower-limit mass has been
found to be $\simeq 1.16 \msol$ for a $\gamma=2$ polytropic EOS, and
$\simeq 0.36 \msol$ for a more realistic model of nuclear matter. It
seems also that these values are independent from the particular
initial velocity profile added to static neutron star models. But,
as it has been noted in Sec.~\ref{ss:IC}, for the polytropic EOS, {\bf
like all masses} this lower-limit mass scales like $K^{0.5}$.

For both studied EOS, in the
case when the star collapses to a black hole it may either proceed
directly, as if it were an {\em unstable} neutron star, all the matter
ending in the black hole; or there may appear a shock and bounce,
ejecting a part of the matter to infinity, so that only a fraction of
the initial neutron star forms a black hole. With such a mechanism,
the resulting black hole mass can be arbitrarily small if one is
fine-tuning the amount of kinetic energy added to the initial neutron
star. This result is in accordance with works by \cite{Cho93} on
critical collapses, more precisely the mass-scaling relation
(\ref{e:relcrit}) also applies here, at least for black hole masses
not too close to the neutron star critical mass. {\bf The
mass scaling exponent $\alpha$ found for the polytropic EOS is rather
different from that found, with quite different initial conditions, by
\cite{NeiC00} for ultrarelativistic $\gamma=2$ EOS.} 

Nevertheless, the
velocity necessary to achieve the collapse to a black hole of a
typical $1.4 \msol$ neutron star is enormous (see
Fig.~\ref{f:mgvpol}). Even in the case of a smaller $K$, leading to a
smaller maximal mass (see Sec.~\ref{ss:IC}), this velocity always
remains larger than $0.1 c$. It is therefore difficult to imagine any
physical process which could inject such an amount of kinetic energy
to a neutron star. Tidal (anisotropic) effects when the neutron star
is in a binary system with another compact object may result in the
destruction of the star (compressed to a ``pancake'', as ordinary
stars passing near a black hole in \cite{MarLB96}) or in its
oscillations. It has been argued (e.g. by \cite{MatW00}) that in a
binary neutron star system, tidal effects could induce such a
compression (i.e. increase of central density) that the stars could
collapse to black holes before their merging. This is very difficult
to achieve, since in our calculations it happened that during the
collapse the central density could reach a value substantially higher than
the critical density $n_B^{\rm crit}$ (in some cases several times
$n_B^{\rm crit}$), whereas the star would not end in a black hole, but
would rather be dispersed or enter an infinite series of
oscillations. Even in the 
case of the neutron star instability transition in tensor-scalar
theory (see \cite{Nov98}), the velocity is never larger than a few percents
of $c$, therefore quite far from values displayed on
Fig.~\ref{f:mgvpol}. Finally, let us mention the recent interesting
work by \cite{HawC00} 
who studied the critical collapses of {\em boson} stars, where the
interaction with a massless real scalar field can result in a
significant transfer of energy from the field to the star, allowing
for the collapse to a black hole of stable configurations. In the case
of neutron stars the interaction with a gravitational wave seems
unlikely to transfer enough energy to lead to the collapse to a black
hole.  

\begin{acknowledgements}
I am indebted to Jos\'e A. Pons for providing with the realistic
equation of state. I also would like to thank Eric Gourgoulhon for his
help and very fruitful discussions, Jos\'e
M$^{\underline{\mbox{a}}}$ Ib\'a\~{n}ez who suggested this study and
the anonymous referee for his comments.
\end{acknowledgements}

\end{document}